# Effect of Adaptive and Fixed Shared Steering Control on Distracted Driver Behavior

Zheng Wang, Satoshi Suga, Edric John Cruz Nacpil, Bo Yang, and Kimihiko Nakano

*Abstract*—Driver distraction is a well-known cause for traffic collisions worldwide. Studies have indicated that shared steering control, which actively provides haptic guidance torque on the steering wheel, effectively improves the performance of distracted drivers. Recently, adaptive shared steering control based on the physiological status of the driver has been developed, although its effect on distracted driver behavior remains unclear. To this end, a high-fidelity driving simulator experiment was conducted involving 18 participants performing double lane changes. The experimental conditions comprised two driver states: attentive and distracted. Under each condition, evaluations were performed on three types of haptic guidance: none (manual), fixed authority, and adaptive authority based on feedback from the forearm surface electromyography of the driver. Evaluation results indicated that, for both attentive and distracted drivers, haptic guidance with adaptive authority yielded lower driver workload and reduced lane departure risk than manual driving and fixed authority. Moreover, there was a tendency for distracted drivers to reduce grip strength on the steering wheel to follow the haptic guidance with fixed authority, resulting in a relatively shorter double lane change duration.

## I. INTRODUCTION

STEERING a car necessitates continuous real-time visual information from the road ahead. Distractions that inhibit safe and timely driver responses to this information are detrimental to driving safety; thus, technology designed to monitor and assist distracted drivers is a crucial advancement [1, 2]. Although fully autonomous driving is not likely to be realized in the near future, partial automation is becoming more readily available in the form of steering assistance systems and driver attention monitoring [3]. Shared steering control systems assist drivers with curve negotiation and lane changes by producing a proper haptic guidance torque on the steering wheel [4, 5], particularly when a lack of attention to visual information results from driver distraction or fatigue [6, 7]. Decreased performance due to a lack of visual information can be compensated for by haptic guidance [8, 9].

Recently, shared steering control with adaptive authority based on driver status has been investigated to further improve driving performance in comparison to fixed authority [10-12]. Adaptive system design increases the automation authority as driver control authority decreases owing to increased workload or decreased workload capacity [13], and this approach is preferable to steering assistance with fixed and rigid authority [14, 15]. Driver workload can be estimated through physiological measures such as eye movements and surface electromyography (sEMG) signals [7]. For steering tasks, drivers can allocate the control authority with regard to the haptic guidance system by adjusting the arm admittance [4]. Previous research has suggested a relationship between arm admittance and grip strength on the steering wheel [16], as measured by sEMG signals from forearm muscles [17]. Inspired by research on the grip-force-based scheduling of guidance forces [18], our previous study adjusted the level of shared steering control according to driver grip strength on the steering wheel to achieve better driver-automation cooperation performance [11]. We found a reduction in both lane departure risk and driver workload to be associated with adaptive authority compared to fixed authority under normal driving conditions [11].

However, the effect of haptic guidance with adaptive authority on distracted driver behavior remains to be determined. Although adaptive automation relieves the driver of the task of engaging and disengaging the automation, it imposes an additional task of monitoring the time-varying automation level of the adaptive system with the possibility of increased workload [13]. Moreover, several experimental studies have indicated that the steering effort of drivers may be even higher for haptic guidance than for manual driving [4, 5]. This situation may be more complicated when drivers become distracted.

Therefore, this study aims to compare the effects of driver distraction, adaptive authority haptic guidance, fixed authority haptic guidance, and manual driving. It is hypothesized that the adaptive haptic guidance would provide more driving safety and comfort to distracted drivers, whereas fixed haptic guidance could more effectively reduce the duration of a lane change.

This work was supported by Grant-in-Aid for Early-Career Scientists No. 19K20318 and No. 21K17781 from the Japan Society for the Promotion of Science.

Corresponding author: Zheng Wang (e-mail: z-wang@iis.u-tokyo.ac.jp)

Z. Wang, B.Yang, and K. Nakano are with the Institute of Industrial Science, The University of Tokyo, Tokyo 153-8505, Japan (e-mail: z-wang@iis.u-tokyo.ac.jp; b-yang@iis.u-tokyo.ac.jp; knakano@iis.u-tokyo.ac.jp).

S. Suga is with the Department of Mechatronics Engineering, Technical University of Darmstadt, 64289 Darmstadt, Germany (e-mail: satosuga95@gmail.com)

E.J.C. Nacpil is with Corpy & Co., Inc., Tokyo 113-0033, Japan (e-mail: edric@corpy.co.jp)

## II. METHOD

### A. Participants

Eighteen healthy subjects (two women and 16 men) were recruited to participate in the experiment. Their ages ranged from 21 to 32 years (mean = 23.5 years, SD = 2.8 years). All subjects had valid Japanese driver's licenses with some driving experience (mean = 2.7 years, SD = 2.6 years). The experiment was approved by the Office for Life Science Research Ethics and Safety, Graduate School of Interdisciplinary Information Studies, University of Tokyo. Each subject provided written consent to the experimental protocol and received compensation for their participation.

### B. Apparatus

A Myo armband (Thalmic Labs, Inc.) acquired the sEMG from the dominant forearm of the driver (Fig. 1). The sEMG was conditioned for further processing by calculating the root mean square (RMS) value of the activation signal from the stainless-steel armband sensors [19]. Driver grip strength was normalized with respect to the maximum sEMG for each participant ($sEMG_{REF}$). The steering wheel torque provided by the adaptive haptic guidance was applied according to the normalized sEMG value.

A high-fidelity moving platform driving simulator with six degrees of freedom was used to run driving scenarios with a projection screen displaying a 140° field of view. An electronic steering system was connected to the host computer of the driving simulator through a controller area network. The electronic steering system comprised a steering wheel, servo motor, and electronic control unit (ECU). The real-time haptic guidance torque calculated by the host computer served as the input to an ECU. A servomotor was subsequently actuated by the ECU to apply a haptic guidance torque to the steering wheel. The haptic guidance torque was calculated based on a model with two look-ahead points, whereas the magnitude and direction of the haptic guidance torque were determined by comparing the target vehicle trajectory based on the fifth-degree Bezier curve with the actual trajectory of the simulated vehicle [11].

### C. Experimental Conditions and Scenario

The participants drove under six conditions, as shown in Table I. Two driver states were considered: attentive and distracted. For each state, there were three types of haptic guidance: HG-Fixed (haptic guidance with fixed authority), HG-Adaptive (haptic guidance with adaptive authority), and manual steering. Using a 6×6 Latin Square, the sequence in which the experimental conditions were presented to the participants was partially counterbalanced. The three Latin squares partially counterbalanced the within-subject order of the conditions.

TABLE I
EXPERIMENTAL CONDITIONS

| Condition | Driver state | Haptic guidance |
|---|---|---|
| 1 | Attentive | Manual (HG-Non) |
| 2 | Attentive | HG-Fixed |
| 3 | Attentive | HG-Adaptive |
| 4 | Distracted | Manual (HG-Non) |
| 5 | Distracted | HG-Fixed |
| 6 | Distracted | HG-Adaptive |

To induce driver distraction, a challenging secondary task called the paced auditory serial addition task was applied during the entire driving course. The subjects were given a number every 3 s and were asked to add the number they just heard with the number they heard before. The attentive experimental condition was a control session in which the driver was under normal driving conditions.

For manual driving, the overall gain of haptic guidance was set to 0, whereas the overall gain of haptic guidance torque was held at 0.25 for HG-Fixed. The normalized torque was based on 25% of the gain for an automated double lane change (DLC). The gain of the haptic guidance torque decreased linearly from 1 (gain for automated DLC) to 0 for HG-Adaptive, when the driver grip strength increased from 0 to $sEMG_{REF}$. HG-Adaptive reduced the haptic guidance to gain more manual control authority with increased grip strength. Grip strength above $sEMG_{REF}$ during the DLC task prompted the haptic guidance gain to be adjusted to 0.

The DLC task, comprising two stages of lane changing, is illustrated with pylons (Fig. 2). To improve the replicability of the experimental results with respect to steering behavior, a PID controller automatically maintained the simulated vehicle speed at 50 km/h.

### D. Experimental Procedure

Each participant mounted the Myo armband on the dominant forearm. The sEMG normalization was realized by having each participant grip the driving simulator steering wheel in a "ten-to-two" position for 2 s with maximum grip strength. This procedure was repeated thrice with 10 s of rest between each repetition. The mean value across all repetitions for a given participant was used as the reference sEMG value, $sEMG_{REF}$, for normalization.

Each subject performed a practice task prior to the actual experiment so that they were accustomed to operating the simulator. Throughout the practice and experimental trials, drivers were trained to maintain both hands at the ten-and-two steering wheel position. The drivers repeated the DLC task five times for each of the six driving conditions, and thus each driver performed a total of 30 trials for the actual experiment. The subjective task load was measured by having each participant complete a questionnaire after each trial.

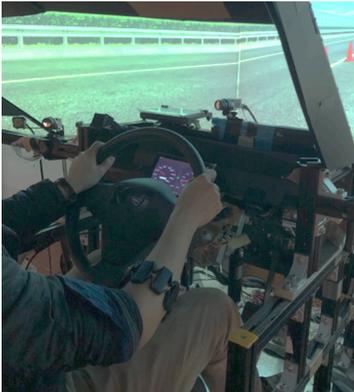

Fig. 1. Active steering system and armband worn by driver to enable sEMG-based operation of high-fidelity driving simulator.

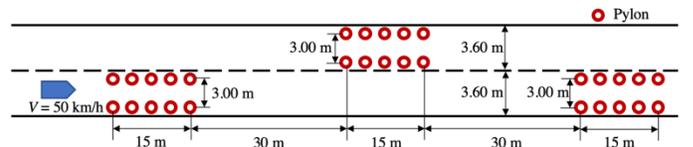

Fig. 2. Experimental driving scenario performed in driving simulator.

*E. Measured Variables*

The DLC performance was evaluated based on driver steering behavior, lane departure risk, and subjective evaluation. The measured driver input torque, DLC duration, steering wheel angle (SWA), and normalized sEMG constituted the evaluation of driver steering behavior. The normalized sEMG RMS ($sEMG/sEMG_{REF}$) was calculated from the measured forearm signals. An increased normalized value indicated greater grip strength. In contrast, the driver steering effort was determined by the calculated RMS value of the driver input torque. Calculating the RMS value of the SWA and peak SWA determined the magnitude of the steering control activity. The peak angle of the SWA of each lane change stage in Fig. 2 was considered, whereas the DLC duration was calculated to quantify the lane changes.

At the conclusion of each lane change stage with the simulated vehicle driving parallel to the entered lane, the lateral error relative to the centerline of the lane was measured. The lane departure risk during DLC was evaluated using the lateral error.

Subjective preferences for the different types of tested haptic guidance paired with each driver state were recorded in conjunction with the NASA task load index (NASA-TLX) to conduct a subjective evaluation. The participants rated their workload according to the index after each driving condition. At the conclusion of all experimental trials, each participant selected the experimental condition with the highest degree of satisfaction. The preference score for a given condition equaled the number of times the condition was selected across all participants. Dividing the preference score by two yielded a relative preference score.

*F. Analysis*

In accordance with two-way repeated-measures analysis of variance (ANOVA), the extent of interaction between the state of the driver and haptic guidance to affect driver behavior was determined. Setting the level of significance to $p = 0.05$, Mauchly's test was executed before the repeated-measures ANOVA. Furthermore, Fisher's least significant difference for pairwise comparisons identified the main effects with a selected significance criterion of $p = 0.05$. Differences were considered statistically significant when the *p*-value was greater than 0.05, and a *p*-value ≤0.1 was interpreted as a tendency toward statistical significance.

### III. RESULTS AND DISCUSSION

The results in this section are described with regard to driver steering behavior and lane departure risk, in addition to subjective evaluation. The values for two-way repeated-measures ANOVA related to driver behavior as well as the corresponding mean and standard deviations for experimental variables are listed in Table II.

The main effect was significant for HG in terms of RMS of driver input torque ($p < 0.001$), lateral error at the end of the first lane change ($p < 0.001$), and relative score of pairwise preference ($p < 0.01$). In contrast, significant differences were observed for HG based on the RMS of SWA ($p < 0.1$) and the RMS of normalized sEMG ($p < 0.1$), although there was no significance for the state of the driver and interaction effect. As for the peak value of SWA in the first lane change and DLC

TABLE II
DEPENDENT VARIABLES OF DRIVER BEHAVIOR AND TWO-WAY REPEATED-MEASURES ANOVA RESULTS

| Variable | Attentive & Manual (1) M (SD) | Attentive & HG-Fixed (2) M (SD) | Attentive & HG-Adaptive (3) M (SD) | Distracted & Manual (4) M (SD) | Distracted & HG-Fixed (5) M (SD) | Distracted & HG-Adaptive (6) M (SD) | Driver State *p* value | HG *p* value | Interaction *p* value |
|---|---|---|---|---|---|---|---|---|---|
| RMS of driver input torque (N·m) | 1.029 (0.064) | 0.734 (0.072) | 0.633 (0.106) | 1.029 (0.058) | 0.722 (0.081) | 0.643 (0.149) | 0.941 | 0.000*** | 0.454 |
| RMS of SWA (deg) | 19.496 (3.998) | 19.740 (3.605) | 20.163 (4.137) | 18.760 (3.860) | 19.580 (4.642) | 19.701 (5.254) | 0.354 | 0.074+ | 0.546 |
| Peak value of SWA at the 1st LC (deg) | 30.406 (10.643) | 30.044 (10.874) | 31.362 (10.771) | 27.409 (9.998) | 28.994 (11.924) | 30.930 (12.372) | 0.172 | 0.013* | 0.083+ |
| Peak value of SWA at the 2nd LC (deg) | -34.728 (7.239) | -34.313 (7.090) | -34.144 (7.787) | -34.126 (6.538) | -33.014 (7.868) | -33.286 (9.580) | 0.282 | 0.567 | 0.825 |
| Duration of double LC (s) | 8.613 (1.044) | 8.437 (0.689) | 8.123 (0.595) | 8.731 (0.788) | 8.088 (0.626) | 8.290 (0.811) | 0.867 | 0.002** | 0.083+ |
| RMS of normalized sEMG (%) | 7.116 (3.501) | 6.746 (3.686) | 6.696 (3.547) | 6.947 (3.966) | 6.309 (3.908) | 7.120 (4.053) | 0.822 | 0.099+ | 0.208 |
| Lateral error at the end of 1st LC (m) | 0.432 (0.203) | 0.354 (0.172) | 0.366 (0.181) | 0.430 (0.204) | 0.364 (0.195) | 0.319 (0.171) | 0.391 | 0.000*** | 0.296 |
| Lateral error at the end of 2nd LC (m) | 0.193 (0.111) | 0.214 (0.123) | 0.169 (0.100) | 0.241 (0.099) | 0.215 (0.101) | 0.224 (0.133) | 0.033* | 0.650 | 0.114 |
| NASA-TLX overall workload | 43.685 (16.020) | 43.019 (15.408) | 40.593 (16.489) | 71.407 (17.924) | 68.130 (17.330) | 66.241 (16.541) | 0.000*** | 0.166 | 0.826 |
| Relative score of pairwise preference | 0.647 (0.786) | 0.882 (0.781) | 1.471 (0.624) | 0.353 (0.702) | 1.059 (0.748) | 1.588 (0.507) | 1 | 0.001** | 0.314 |

+ $p < 0.1$, * $p < 0.05$, ** $p < 0.01$, *** $p < 0.001$
SWA: Steering wheel angle; LC: Lane change; HG: Haptic guidance.

duration, the main effect was significant for HG (p < 0.05, p < 0.01), and there was a tendency toward significance for the interaction effect (p < 0.1, p < 0.1). However, there were no significant differences in the state of the driver. In terms of lateral error at the end of the second lane change and overall workload according to the NASA-TLX, the main effect was significant for the state of the driver (p < 0.05, p < 0.001), whereas no significant difference was observed for the HG and the interaction effect.

*A. Driver Steering Behavior*

As shown in Table II, pairwise comparisons indicated that the driver input torque for manual was significantly greater than that for HG-Fixed (p < 0.001), higher for Manual than for HG-Adaptive (p < 0.001), and greater for HG-Fixed than for HG-Adaptive (p < 0.001). Therefore, our hypothesis was validated because haptic guidance significantly reduced driver steering effort, and HG-Adaptive was more effective.

The peak values of SWA in the first lane change are shown in Table II and Fig. 3. For the distracted state, the peak value of SWA was significantly lower for manual steering than for HG-Adaptive (p < 0.01), lower for HG-Fixed than for HG-Adaptive (p < 0.05), and lower for Manual than for HG-Fixed (p < 0.1). Moreover, the peak value of SWA for manual steering was significantly higher with Attentive than with Distracted (p < 0.05). Hence, the distraction reduced the steering activity, and the haptic guidance system increased the steering activity to a level comparable to that of the Attentive state. A similar tendency can be found for the RMS of the SWA, although the difference was not statistically significant. This may be due to the effect of distraction on SWA not being pronounced during the second lane change, as shown in Table II. Thus, a more demanding secondary task or lane change task will be considered in future studies.

The results for the DLC duration are shown in Table II and Fig. 4. From pairwise comparisons, for Attentive, the duration of the DLC was significantly shorter with HG-Adaptive than with HG-Fixed with p < 0.05, whereas the DLC tended to be significantly shorter for HG-Adaptive than for manual steering, where p < 0.1. For Distracted, the DLC duration was significantly shorter with HG-Fixed than with Manual (p < 0.001) and shorter with HG-Adaptive than with Manual (p < 0.01). Moreover, for HG-Fixed, there was a tendency for the DLC duration to be shorter with Distracted than with Attentive (p < 0.1). Therefore, HG-Fixed could more effectively reduce lane change duration for distracted drivers, as predicted by our hypothesis.

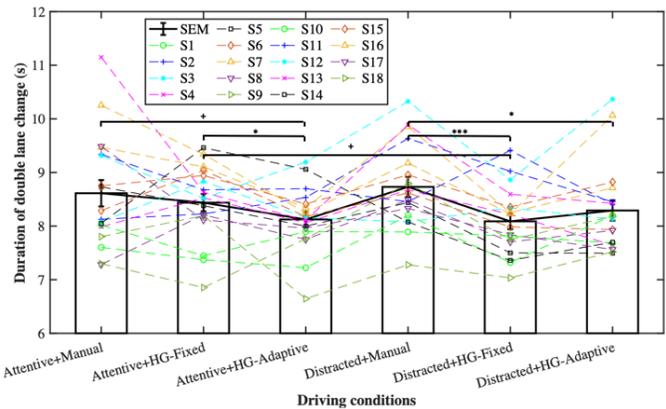

Fig. 4. Double lane-change duration. Mean +/- SEM represented by error bars.

The RMS results of sEMG are shown in Table II and Fig. 5. Based on pairwise comparisons, for Distracted, the RMS of sEMG was significantly lower with HG-Fixed than with HG-Adaptive (p < 0.05) and lower with HG-Fixed than with Manual (p < 0.1), indicating that distracted drivers tended to give more control authority to the HG-Fixed by reducing grip strength. Consequently, the DLC duration was relatively shorter (Fig. 4).

*B. Lane Departure Risk*

Fig. 6 and Table II show the results of the lateral error at the end of the first lane change. Pairwise comparisons indicated that for the Attentive condition, the lateral error was significantly higher for manual steering than for HG-Fixed with p < 0.05 and tended to be higher for HG-Adaptive with p < 0.1. The lateral error for Distracted was significantly lower in the case of HG-Fixed than in Manual, where p < 0.05, HG-Adaptive was significantly lower than with HG-Fixed (p < 0.01). Furthermore, HG-Adaptive tended to be significantly lower than HG-Fixed with p < 0.1. Thus, haptic guidance can reduce lane departure risk when the driver is attentive, as confirmed by our previous study [11]. This outcome could be attributed to the human driver being limited by the response of the neuromuscular system, thereby making it difficult to complete the DLC accurately [20]. Furthermore, from this result, haptic guidance is also capable of reducing the lateral error for distracted drivers, with HG-Adaptive being more effective.

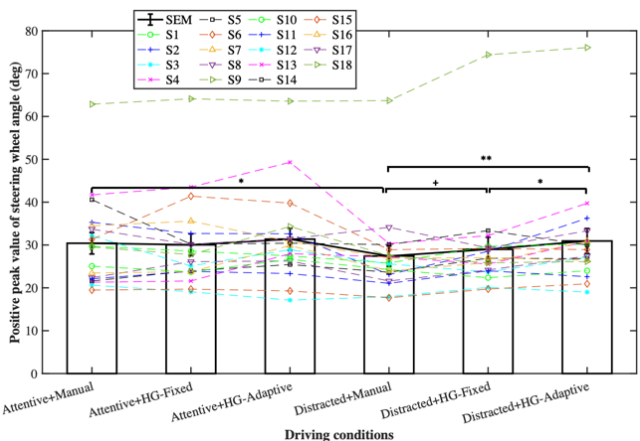

Fig. 3. Peak value of SWA at the first lane change. Mean +/- SEM (standard error of mean) represented by error bars.

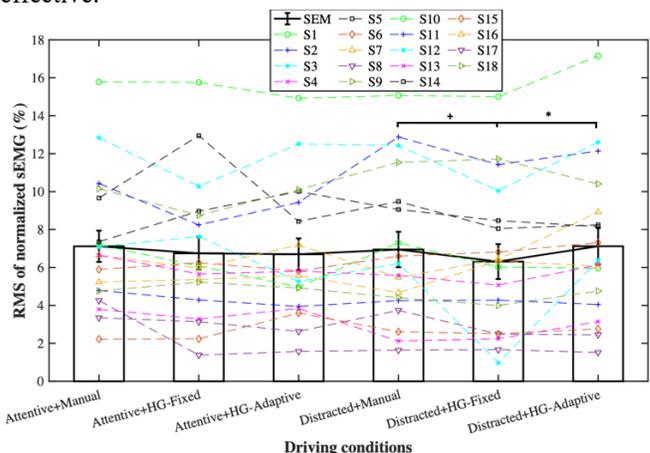

Fig. 5. RMS of normalized sEMG (%). Mean +/- SEM represented by error bars

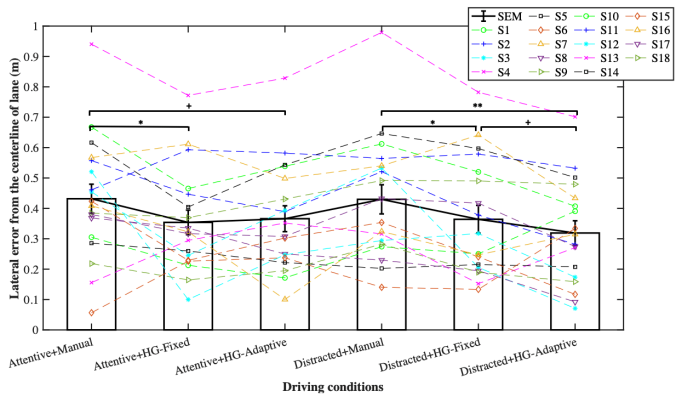

Fig. 6. Lateral error with respected to lane centerline at end of first lane change. Mean +/- SEM represented by error bars.

The results for lateral error in the case of the second lane change are listed in Table II. The lateral error was significantly increased by driver distraction, but the effect of haptic guidance was insignificant. According to the peak SWA in the second lane change, no significance was observed with regard to the effect of HG. Thus, in accordance with our previous study [11], the proposed haptic guidance system is more effective for the first lane change than the second one. This may be because the lateral error at the first lane change is almost twice as large as the second lane change (shown in Table II) because of more frequent overshoot, and haptic guidance is effective in diminishing the overshoot.

### C. Subjective Evaluation

The results of the driver workload assessed by the NASA-TLX are shown in Table II and Fig. 7. Taking into account the pairwise comparisons, the overall driver workload for Distracted was significantly higher than for Attentive ($p < 0.001$). Moreover, for Distracted, HG-Fixed yielded a lower overall workload ($p < 0.1$), lower physical demand ($p < 0.1$), and lower effort ($p < 0.1$), and HG-Adaptive yielded a lower temporal demand ($p < 0.1$). This indicates that the driver workload was increased by the secondary task, and the haptic guidance system effectively reduced the driver workload.

Table II shows the relative scores of pairwise driver preferences. For both Attentive and Distracted, drivers preferred HG-Fixed over Manual, HG-Adaptive over Manual, and HG-Adaptive over HG-Fixed. Moreover, there is a tendency for more drivers to prefer haptic guidance over the manual when they are distracted, which is expected, as haptic guidance reduces lane departure risk as well as driver workload.

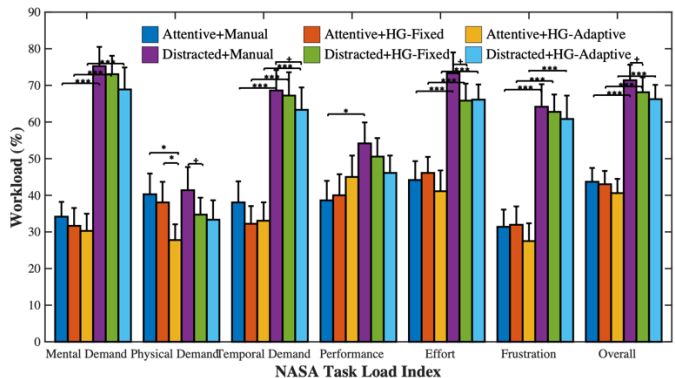

Fig. 7. Mean scores on NASA-TLX. Data error bars represent mean + SEM.

## IV. Conclusion

This driving simulator study focused on the effect of haptic guidance with adaptive authority on distracted driver behavior. DLC tasks were completed according to experimental conditions that were designed by combining two driver states, namely Attentive and Distracted, along with three haptic guidance categories: HG-Fixed, HG-Adaptive, and Manual. HG-Adaptive relied on feedback from the real-time forearm sEMG of the driver.

For both attentive and distracted drivers, HG-Adaptive yielded a greater reduction in the driver workload and lane departure risk than that of HG-Fixed and manual steering. Moreover, drivers tended to reduce the steering wheel grip strength to provide admittance to the haptic guidance with fixed authority, resulting in the completion of the DLC in a relatively short period. As the current study indicated a small lane departure risk induced by driver distraction, we plan to address the possibility of greater risk by conducting a future study with a more demanding secondary task or lane change task.

## References

[1] A. Koesdwiady, R. Soua, F. Karray, and M. S. Kamel, "Recent trends in driver safety monitoring systems: State of the art and challenges," *IEEE Trans. on Veh. Technol.*, vol. 66, no. 6, pp. 4550-4563, Jun. 2017.

[2] Y. Xing, C. Lv, H. Wang, D. Cao, E. Velenis, and F.-Y. Wang, "Driver activity recognition for intelligent vehicles: A deep learning approach," *IEEE Trans. on Veh. Technol.*, vol. 68, no. 6, pp. 5379-5390, Jun. 2019.

[3] M. J. Jensen, A. M. Tolbert, J. R. Wagner, F. S. Switzer, and J. W. Finn, "A customizable automotive steering system with a haptic feedback control strategy for obstacle avoidance notification," *IEEE Trans. on Veh. Technol.*, vol. 60, no. 9, pp. 4208-4216, Nov. 2011.

[4] D. A. Abbink, M. Mulder, and E. R. Boer, "Haptic shared control: smoothly shifting control authority," *Cogn., Technol. Work*, vol. 14, no. 1, pp. 19-28, Mar. 2012

[5] K. K. Tsoi, M. Mulder, and D. A. Abbink, "Balancing safety and support: Changing lanes with a haptic lane-keeping support system," in *Proc. IEEE Int. Conf. Syst., Man, Cybern.*, Istanbul, Turkey, Oct. 2010, pp. 1236-1243.

[6] P. G. Griffiths and R. B. Gillespie, "Sharing control between humans and automation using haptic interface: Primary and secondary task performance benefits," *Human Factors, J. Human Factors Ergonom. Soc.*, vol. 47, no. 3, pp. 574-590, Sep. 2005.

[7] Z. Wang, R. Zheng, T. Kaizuka, K. Shimono, and K. Nakano, "The effect of a haptic guidance steering system on fatigue-related driver behavior," *IEEE Trans. Human-Mach. Syst.*, vol. 47, no. 5, pp. 741-748, Oct. 2017.

[8] F. Mars, M. Deroo, and J.-M. Hoc, "Analysis of human-machine cooperation when driving with different degrees of haptic shared control," *IEEE Trans. Haptics*, vol. 7, no. 3, pp. 324-333, Jul./Sep. 2014.

[9] Z. Wang, R. Zheng, T. Kaizuka, and K. Nakano, "Influence of haptic guidance on driving behaviour under degraded visual feedback conditions," *IET Intell. Transp. Syst.*, vol. 12, no. 6, pp. 454-462, Aug. 2018.

[10] X. Ji, K. Yang, X. Na, C. Lv, and Y. Liu, "Shared steering torque control for lane change assistance: A stochastic game-theoretic approach," *IEEE Trans. on Ind. Electron.*, vol. 66, no. 4, pp. 3093-3105, Apr. 2019.

[11] Z. Wang, S. Suga, E. J. C. Nacpil, Z. Yan, and K. Nakano, "Adaptive driver-automation shared steering control via forearm surface electromyography measurement," *IEEE Sensors J.*, vol. 21, no. 4, pp. 5444-5454, Feb. 2021.

[12] M. Li, H. Cao, X. Song, Y. Huang, J. Wang, and Z. Huang, "Shared control driver assistance system based on driving intention and situation assessment," *IEEE Trans. on Ind. Inform.*, vol. 14, no. 11, pp. 4982-4994, Nov. 2018.

[13] D. B. Kaber, J. M. Riley, K.-W. Tan, and M. R. Endsley, "On the design of adaptive automation for complex systems," *Int. J. Cogn. Ergonom.*, vol. 5, no. 1, pp. 37-57, Jun. 2001.

[14] B. Zhu, S. Yan, J. Zhao, and W. Deng, "Personalized lane-change assistance system with driver behavior identification," *IEEE Trans. on Veh. Technol.*, vol. 67, no. 11, pp. 10293-10306, Nov. 2018.

[15] Y. Jiang, W. Deng, J. Wu, S. Zhang, and H. Jiang, "Adaptive steering feedback torque design and control for driver–vehicle system considering driver handling properties," *IEEE Trans. on Veh. Technol.*, vol. 68, no. 6, pp. 5391-5406, Jun. 2019.

[16] H. Nakamura, D. Abbink, and M. Mulder, "Is grip strength related to neuromuscular admittance during steering wheel control," in *Proc. IEEE Int. Conf. Syst., Man, Cybern.*, Anchorage, AK, USA, Oct. 2011, pp. 1658-1663.

[17] Z. Wang, T. Kaizuka, K. Nakano, R. Zheng, and K. Shimono, "Evaluation of driver steering performance with haptic guidance under passive fatigued situation," in *Proc. IEEE Int. Conf. Syst., Man, Cybern.*, Budapest, Hungary, Oct. 2016, pp. 3334-3339.

[18] J. Smisek, W. Mugge, J. B. Smeets, M. M. van Paassen, and A. Schiele, "Haptic guidance on demand: A grip-force based scheduling of guidance forces," *IEEE Trans. Haptics*, vol. 11, no. 2, pp. 255-266, Apr./Jun. 2018.

[19] M. Tomaszewski. GitHub. *Myo SDK MATLAB MEX Wrapper*. Accessed: Jan. 7, 2020. [Online]. Available: https://www.github.com/mark- toma/MyoMex

[20] G. Chen, S. Chen, R. Langari, X. Li, and W. Zhang, "Driver-behavior-based adaptive steering robust nonlinear control of unmanned driving robotic vehicle with modeling uncertainties and disturbance observer," *IEEE Trans. on Veh. Technol.*, vol. 68, no. 8, pp. 8183-8190, Aug. 2019.